\documentclass[aps,prl,twocolumn,showpacs,preprintnumbers,amsmath,amssymb]{revtex4}

\usepackage{graphicx}
\usepackage{dcolumn}
\usepackage{bm}

\begin{document}

\preprint{}

\title{Analytical calculation of the drag force near drag crisis of a
 falling sphere\\}

\author{Armando V.D.B. Assis}
\affiliation{Departamento de F\'isica,
Universidade Federal de Santa Catarina, 
88040-900, Florian\'opolis, SC, Brazil\\}
\author{M.H.R. Tragtenberg}
\affiliation{Departamento de F\'isica,
Universidade Federal de Santa Catarina, 
88040-900, Florian\'opolis, SC, Brazil\\}
\author{N.S. Branco}
\affiliation{Departamento de F\'isica,
Universidade Federal de Santa Catarina, 
88040-900, Florian\'opolis, SC, Brazil\\}


\date{\today}

\begin{abstract}
We obtain analitically the $v^2$ dependence of the drag force on a falling sphere close to the drag crisis, as well as the drag coefficient at the drag crisis, with excellent agreement with experiment. We take into account the effects of viscosity in creating a turbulent boundary layer and perform the calculations using the Navier-Stokes equation.\end{abstract}

\maketitle

We report in this paper the first derivation of both the $v^2$ dependence of the drag force and the drag coefficient at the drag crisis (abrupt decrease of the drag coefficient) for a falling sphere, which is in excellent agreement with experimental results and is derived from the Navier-Stokes equation. Previous works are far from the well-known experimental results \cite{ref2,ref13}. We remark that this is, to our knowledge, the first analytical derivation taking into account the Navier-Stokes equation and the character of the boundary layer, and which is in excellent agreement with experimental data.

There is no theoretical determination of the
velocity dependence of the drag force based on the Navier-Stokes equation that take into account the physics of the boundary layer and accomplish with the dependence of the drag force on the sphere velocity close to the drag crisis. We understand that these achievements are important steps towards the full understanding of the underlying aspects of the physics of the drag crisis. The works in the literature that 
support our results are \cite{ref1,ref2,ref3,ref4,ref5,ref6,ref7,ref8,ref9,ref10,ref11,ref12,ref13}, in agreement with the result for the drag coefficient reported in this  paper. 

We start considering the case of the sphere falling in the stationary state, with constant velocity in relation to the ground. We consider the reference frame fixed on the ground, where the Navier-Stokes equation for an isotropic fluid reads 
\begin{equation}
\rho\dot{\vec{v}}-\rho\vec{g}+\vec{\nabla}p-\vec{\nabla}\cdot\Gamma=\vec{0}\label{1},
\end{equation}
\begin{equation}
\Gamma_{ik} = \eta\left(\frac{\partial v_{i}}{\partial x_{k}} + \frac{\partial v_{k}}{\partial x_{i}} 
- \frac{2}{3}\delta_{ik}\frac{\partial v_{\lambda}}{\partial x_{\lambda}}\right) + 
\zeta\delta_{ik}\frac{\partial v_{\lambda}}{\partial x_{\lambda}}\label{2},
\end{equation}\\ 
where $\vec{v}$ is the fluid velocity field,  $\rho$ is the fluid density, $\vec{g}$ is the external gravitational field, $m$ is the sphere mass, $p$ is the pressure field, $\eta$ and $\zeta$ are the first and the second viscosity coefficients taken as constants and the dot denotes the total time derivative. The Einstein summation convention is being used on repeated indices. For divergence free velocity field fluids (constant density is a sufficient condition), the Navier-Stokes equations turns out to be 
\begin{equation}
\rho\dot{\vec{v}}-\rho\vec{g}+\vec{\nabla}p-\eta\vec{\nabla}^{2}\vec{v}=\vec{0}\label{4}. 
\end{equation}
When we go to the reference frame fixed on the sphere, the Navier-Stokes has the same form, but now $\vec{g}\rightarrow-\vec{F}/m$, where $\vec{F}$ is the fluid force on the sphere at rest. We assume null fluid velocity at the sphere surface, the nonslip boundary condition on the surface of the sphere, and steady state fluid velocity $-\dot{h}(t)\hat{e}_{z}$ far away from the sphere, where $\dot{h}(t)\hat{e}_{z}$ is the the fall velocity of the center of the sphere in relation to the ground and $\hat{e}_{z}$ is the vertical upwards versor.

We want to determine the force $\vec{F}$ the fluid exerts on the sphere. Back to the general case, from the continuity and Navier-Stokes equations we arrive to the equation below for each component: 
\begin{equation}
\frac{\partial}{\partial t}\left(\rho v_{j}\right)=-\frac{\rho F_{j}}{m}+\frac{\partial}{\partial x_{k}}
\left(-\delta_{jk}p+\Gamma_{jk}-\rho v_{j} v_{k}\right)\label{9}.
\end{equation}
Integrating Eq.4 in a fixed, arbitrary and nondeformable control volume, we have:
\begin{equation}
\frac{\partial}{\partial t}\int\rho\vec{v}dV=-\int\rho\frac{\vec{F}}{m}dV+\oint\Pi\cdot\hat{n}dS,
\end{equation}
where $\Pi$ is the tensor:
\begin{equation}
\Pi=-\mathbf{1}p+\Gamma-\rho\left(\vec{v}\otimes\vec{v}\right),
\end{equation}
$\mathbf{1}$ is the identity tensor and $\vec{v}\otimes\vec{v}$ is the dyadic product.
In general case, $-\vec{F}/m$ is the non-inertial acceleration field, the same at any point of the non-inertial frame fixed on the sphere at a given instant. Hence, the fluid force on the sphere is given by
\begin{equation}
\frac{\vec{F}}{m}=\frac{1}{\int_{CV}\rho dV}\left(\oint_{CS}\Pi\cdot\hat{n}dS-
\frac{\partial}{\partial t}\int_{CV}\rho\vec{v}dV\right),
\end{equation}\\
\vspace{-0.4cm}
\\
where CV and CS are the control volume and the control surface, respectively. 

The fluid velocity field may be divided in three main regions: the mainstream, the boundary layer and the wake in Fig. 1. 
\begin{center}
\includegraphics[scale=0.35]{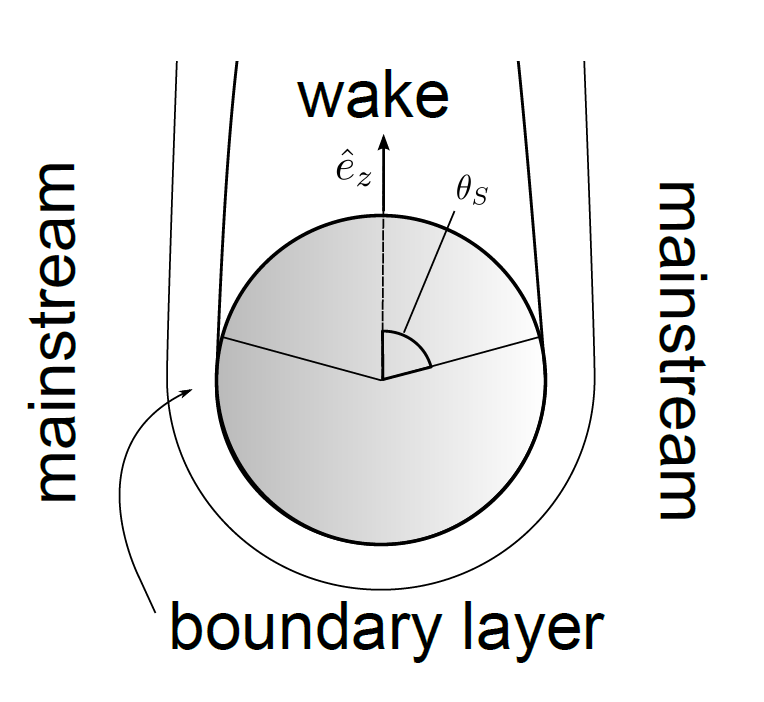}\\
Figure 1: Schematic draw of the main regions.
\end{center}
The stationary velocity field in the mainstream $\vec{v}_{\infty}(\vec{r})$ is taken as irrotational in virtue of the rapid decrease of deformation outside the boundary layer compared to the internal boundary layer region, at high Reynolds numbers. Hence, the velocity field is taken as potential flow  $\vec{v}_{\infty}(\vec{r})=\vec{\nabla}K(\vec{r})$ outside the boundary layer. For an incompressible fluid, $K(\vec{r})$ must obey the Laplace equation. The boundary conditions for the stationary velocity field are: at infinity it is $\displaystyle\lim_{\left|\vec{r}\right| \to \infty}\vec{v}_{\infty}(\vec{r})=
-\dot{h}^{\infty}(t)\hat{e}_{z}$,  whereas it is tangent to the outer surface of the boundary layer. 
The solution for the 
stationary fluid velocity field at the mainstrean is 
\begin{eqnarray}
\vec{v}_{\infty}(\vec{r}) & = & -\dot{h}^{\infty}(t)\left[1-\frac{\left(R+\delta\right)^{3}}{r^{3}}\right]\cos\theta\hat{e}_{r}\nonumber \\
                          &   &+\dot{h}^{\infty}(t)\left[1+\frac{\left(R+\delta\right)^{3}}{2r^{3}}\right]\sin\theta\hat{e}_{\theta},
\end{eqnarray}\\
where R is the sphere radius and $\delta$ is the thickness of the boundary layer.
The mainstream stationary velocity field obeys the Navier-Stokes equation in the form of Eq. (3) (with $\vec{g}\rightarrow-\vec{F}/m$). Since $\vec{\nabla}\times\vec{F}=\vec{0}$, we define a potential $\varphi^\infty$. Thus, we find a Bernoulli field, such that:
\begin{equation}
p_{\infty}=p_{\infty}^{0}-\frac{\rho}{m}\varphi^{\infty}-\rho\frac{v_{\infty}^{2}}{2},
\end{equation}
where $p_{\infty}$ is the pressure scalar field in the mainstream region, $p_{\infty}^{0}$ a constant pressure field.
In order to evaluate the force $\vec{F}^{\infty}$ the fluid exerts on the sphere we recall the Eqs. (6), (7), in the stationary state
\begin{equation}
\vec{F}^{\infty}=\frac{m}{\int_{CV}\rho dV}\left(\oint_{CS}\Pi^{\infty}\cdot\hat{n}dS-\frac{\partial}{\partial t}\int_{CV}\rho\vec{v}_{\infty}dV\right),
\end{equation}
\begin{equation}
\Pi^{\infty}=\left.\left[-\mathbf{1}p_{\infty}+\Gamma_{\infty}-
\rho\left(\vec{v}_{\infty}\otimes\vec{v}_{\infty}\right)\right]\right|_{CS}.
\end{equation}
We need to define an appropriate control volume and its associated control surface. We choose
the control surface as the AFGBA surface, where BA is in contact with the wake, as can be 
seen in Fig. 2.
\begin{center}
\includegraphics[scale=0.3]{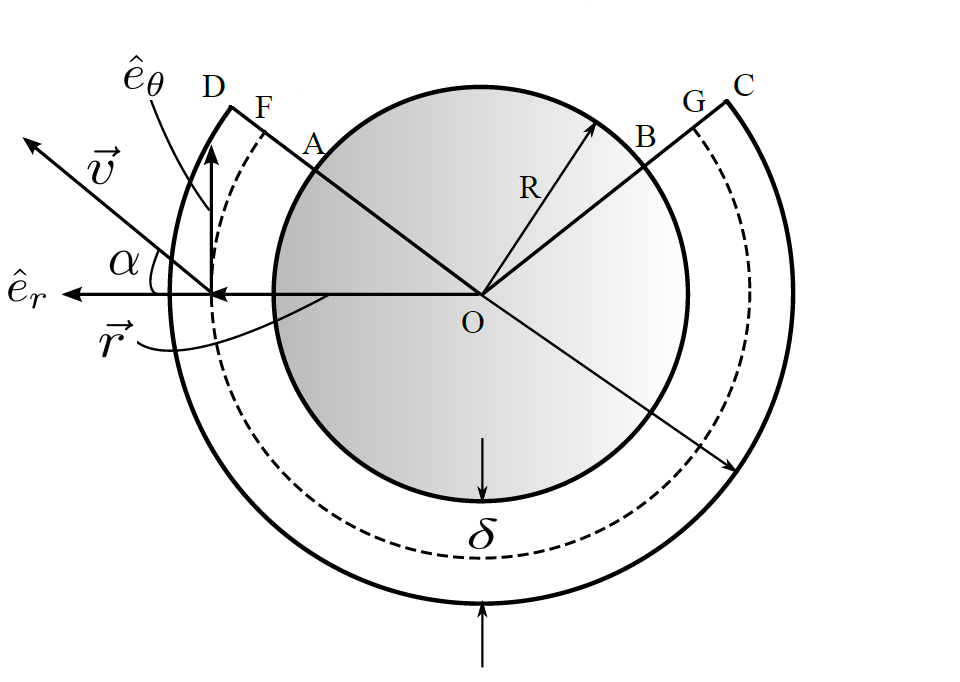}\\
Figure 2: Schematic draw of the control region.
\end{center}
The time dependent term in the rigth hand side of Eq. (10) can be determined under the assumption that we are interested in the brink of the drag crisis, where the boundary layer is in a turbulent transition regime. In that regime, the turbulent boundary layer does not have a stationary state. But we are interested in the average effect, which can be taken into account by considering the ensemble average, which is time independent. Then, we can neglect this term.

We can also neglect the term $\Gamma_{\infty}$ of Eq. (11), considering that the turbulent boundary layer has a bigger velocity gradient in the surface AB (AB in contact with the boundary layer) in comparison with the velocity gradient in FG (the onset profile of the velocity internal to the boundary layer will be, in few lines below, modeled by a step function). This is because we are close to the drag crisis, and the nonslip boundary condition imposes big changes in velocity from the surface to regions close to the sphere, within the boundary layer, and we are able to choose the surface FG as far as we want from the sphere surface, in order to have a smaller velocity gradient, even though this gradient is within a turbulent region.

The integral in the denominator of Eq. (10) is $m+m_{BL}$, the sum of the mass of the sphere with the mass of the boundary layer. The tensor $\vec{v}_{\infty}\otimes\vec{v}_{\infty}$ is null at the surface AB in contact with the wake, by the non-slip boundary condition, as well as at the surface close to the separation points (AF and GB in Fig. 2), since the velocity is close to zero in this region. We end up with
\begin{widetext}
\begin{equation}
\vec{F}^{\infty}=\frac{m}{m+m_{BL}}\Bigg\{-\left[\int_{FG}+\int_{GB}+\int_{BA}+\int_{AF}\right]\mathbf{1}p_{\infty}\cdot\hat{n}dS-\int_{FG}\rho\left(\vec{v}_{\infty}\otimes\vec{v}_{\infty}\right)\cdot\hat{n}dS\Bigg\}.
\end{equation}
\end{widetext}
In order to calculate $\vec{F}^{\infty}$, we have to determine the steady state pressure field and the tensor $\left(\vec{v}_{\infty}\otimes\vec{v}_{\infty}\right)$. Concerning the pressure field for a laminar boundary layer, we have no change in this field for different distances from the surface of the sphere
\begin{equation}
\frac{\partial p}{\partial r}\approx 0.
\end{equation}
\cite{ref2}. However, for a turbulent boundary layer, close to the drag crisis, Eq. (13) is not true for the pressure field $p$, but it is a good approximation for the pressure field $p^{'}=p+\rho\varphi_{\vec{F}}/m$. This pressure field incorporates the potential related to the non-inertial irrotational force field. We can assume this in the region between the surfaces FG and CD, since this region is sufficiently thin and laminar. Then,
\begin{equation}
p_{\infty}^{FG} =-\frac{\rho}{m}\varphi^{\infty}_{FG}+p^{0}_{\infty}-\frac{9}{8}\rho\left[\dot{h}^{\infty}(t)\right]^{2}\sin^{2}{\theta}.
\end{equation}

The pressure field in the GBAF surface (BA in contact with the wake) will be constant \cite{ref1,ref8}. Then, if we are able to calculate the pressure field at the separation point S ($\equiv$ B), we can calculate it in that surface. We invoke that we can draw an SG ($\equiv$ BG) current line, \textit{at the drag crisis}, close to the B and G points
where we have low shear, since it is the separation point neighborhood. Throughout this line, Euler equation is valid and Bernoulli law too. Then we are able to support that
\begin{equation}
p^{S}_{\infty}=p_{\infty}^{G}+\frac{1}{2}\rho v_{\infty}^{2}\left(R+\delta',\,\theta_{S}\right),
\end{equation}
since $\delta^{'}<<R$, where $\delta^{'}$ is the radial distance from the sphere surface until the 
FG surface. Then the pressure field close to wake and the separation point, is given by
\begin{widetext}
\begin{equation}
p_{\infty}^{S}=-\frac{\rho}{m}\varphi^{\infty}\left(R+\delta',\,\theta_{S}\right)+p^{0}_{\infty}-\frac{9}{8}\rho\left[\dot{h}^{\infty}(t)\right]^{2}\sin^{2}{\theta_{S}}+\frac{1}{2}\rho v_{\infty}^{2}\left(R+\delta',\,\theta_{S}\right).
\end{equation}
\end{widetext}

We will model the square velocity profile inside the boundary layer as a step function, since it is a high gradient velocity field close to the drag crisis, and we are able to expand the time average of this field into a Fourier series. The value of the square velocity at the distance $\delta'$ is then given by
\begin{equation}
v_{\infty}^{2}\left(R+\delta',\,\theta\right)=\frac{9}{8}\left[\dot{h}^{\infty}(t)\right]^{2}\sin^{2}{\theta}.
\end{equation} 
Then, the pressure field close to the separation point will be
\begin{equation}
p_{\infty}^{S}= -\frac{\rho}{m}\varphi^{\infty}_{GBAF}+p^{0}_{\infty}-\frac{9}{16}\rho\left[\dot{h}^{\infty}(t)\right]^{2}\sin^{2}{\theta_{S}},
\end{equation}
where GBAF is the region composed by the wake in contact with the sphere
plus the neighboring region to the separation point. Since the region FG is
turbulent, the velocity field is not tangential there. We have to perform
a time average there. The turbulent velocity field, there, is 
\begin{equation}
\vec{v}_{FG}=v_{FG}\left(R+\delta',\theta,t\right)\left[\cos{\alpha(t)}\,\hat{e}_{r}+\sin{\alpha(t)}\,\hat{e}_{\theta}\right],
\end{equation}
where $\alpha$ is the angle between the velocity of the fluid and the normal
vector of the FG spherical surface (see Fig. 2).
Then, the tensor which gives the linear momentum flux through FG is $\left(\vec{v}\otimes\vec{v}\right)_{FG}\cdot\hat{n}=\vec{v}_{FG}\left(R+\delta',\theta,t\right)\left[\vec{v}_{FG}\left(R+\delta',\theta,t\right)\cdot\hat{n}\right]$, and has
an \textit{ensemble} average given by
\begin{equation}
\left\langle v_{FG}^{2}\left(R+\delta',\theta,t\right)\left[\cos^{2}{\alpha(t)}\,\hat{e}_{r}+\cos{\alpha(t)}\sin{\alpha(t)}\,\hat{e}_{\theta}\right]\right\rangle_{t}.
\end{equation}
Assuming randomness in the angles $\alpha$ between the fluid elements and
the normal of the sphere surface, we find that
\begin{displaymath}
\left(\vec{v}_{\infty}\otimes\vec{v}_{\infty}\right)\cdot\hat{n}=\frac{1}{2}\left\langle v_{FG}^{2}\left(R+\delta',\theta,t\right)\right\rangle_{t}\hat{e}_{r}
\end{displaymath}
\vspace{-0.7cm}
\begin{equation}
\,\,\,\,\,\,\,\,\,\,\,\,\,\,\,\,\,\,\,\,\,\,\,\,\,\,\,\,\,=\frac{9}{16}\left[\dot{h}^{\infty}(t)\right]^{2}\sin^{2}{\theta}\,\hat{e}_{r}.
\end{equation}
From the Eqs. (14), (18), (21) and (12) we are able to calculate the force the fluid
exerts on the sphere
\begin{widetext}
\begin{equation}
\left(1+\frac{m_{CL}}{m}\right)\vec{F}^{\infty}=\text{buoyancy force}+\frac{m_{CL}}{m}\vec{F}^{\infty}+\frac{9\pi}{32}\rho\left[\dot{h}^{\infty}(t)\right]^{2} R^{2}\sin^{4}{\theta_{S}}\,\hat{e}_{z}. 
\end{equation}
\end{widetext}
The drag force is eventually given by
\begin{equation}
\vec{F}_{\text{d}}=\frac{9\pi}{32}\rho\left[\dot{h}^{\infty}(t)\right]^{2} R^{2}\sin^{4}{\theta_{S}}\,\hat{e}_{z},
\end{equation}\\
where 
$\vec{F}_{\text{d}}$ is the viscous drag force, since $\vec{F}^{\infty}/m=-\vec{g}$ in the stationary state.
The drag coefficient  $C_D$ at the drag crisis is
\begin{equation}
C_{D}=\frac{2F_{\text{a}}}{\pi\rho R^{2}\left[\dot{h}^{\infty}(t)\right]^{2}}=0.44,
\end{equation}\\
since $\theta_{S}=70,4^{\circ}$ is the separation angle \cite{ref1}.
This coefficient is in excellent agreement with experimental results \cite{ref10}.
\section{Acknowledgments\protect\\} 
The authors would like to thank CNPq and CAPES for partial financial support.
\bibliography{bibliografia}
\end{document}